\documentclass[preprint,12pt]{elsarticle}

\usepackage{amssymb}
\usepackage{tikz}
\usepackage{amsmath}
\usepackage{subcaption}
\usepackage{url}
\usepackage{epstopdf}   

\journal{Journal of Chemometrics}

\begin{document}

\begin{frontmatter}



\title{An alignment-agnostic methodology for the analysis of designed separations data}


\author{Michael Sorochan Armstrong}
\author{Jos\'e Camacho}
\address{Computational Data Science (CoDaS) Lab, Department of Signal Theory, Networks and Communication - University of Granada,
            C/ Periodista Daniel Saucedo Aranda, 
            Granada,
            18071, 
            Andalusia,
            Spain}
\begin{abstract}
Chemical separations data are typically analysed in the time domain using methods that integrate the discrete elution bands. Integrating the same chemical components across several samples must account for retention time drift over the course of an entire experiment as the physical characteristics of the separation are altered through several cycles of use. Failure to consistently integrate the components within a matrix of $M \times N$ samples and variables create artifacts that have a profound effect on the analysis and interpretation of the data. This work presents an alternative where the raw separations data are analysed in the frequency domain to account for the offset of the chromatographic peaks as a matrix of complex Fourier coefficients. We present a generalization of the permutation testing, and visualization steps in ANOVA-Simultaneous Component Analysis (ASCA) to handle complex matrices, and use this method to analyze a synthetic dataset with known significant factors and compare the interpretation of a real dataset via its peak table and frequency domain representations.

\end{abstract}


\begin{highlights}
\item Raw GC-FID signal from a designed experimental dataset is treated with a Fast Fourier Transform (FFT) and analysed using a generalization of ANOVA-Simultaneous Component Analysis (ASCA) to handle complex data.
\item The results of the FFT-ASCA analysis are largely consistent with the analysis using the peak table data. This is confirmed using a demonstration with synthetic data.
\end{highlights}


\end{frontmatter}


\section{Introduction}
\label{sec:sample1}
\subsection{Challenges with peak table analysis}
Chromatographic separations enable detailed analyses of chemical mixtures, providing valuable information for researchers interrogating complex systems. For a multivariate analysis, several samples are collected and discrete chemical bands are integrated within time intervals whose apices are recorded as ``retention times'' - a descriptor of a chemical feature based on the elapsed time between injection and detection of the compound. This is typically done separately for each sample, and issues arise when the signal for each chemical feature is close to the baseline noise, or convolved with other closely-eluting chemical features. Within the multivariate context these challenges do become more significant since a faithful representation of the chemical information must account for the fact that the same chemical features may drift between analytical runs, a phenomenon often referred to as ``peak drift''~\cite{christensen2005chromatographic}. Failure to integrate the same chemical features, or chromatographic peaks in the same column in an $M \times N$ matrix of experimental information (i.e., the ``peak table'') introduces artifacts that affect the resulting inferences from the the analysis, since a peak may be falsely indicated as being distinct from similar chemical features in other samples (a false negative), or incorrectly associated with a chemical feature it is actually different from (similar to a false positive). In the former case, either zeros are introduced to an $n \in \{1...N\}$ column which is easy to notice from an inspection of the data, or possibly summed together with an unrelated feature which is much harder to identify. Both instances break the assumption of bilinearity within the context of a multivariate analysis, since these artifacts make covariance of the data poorly representative of the actual chemical phenomena being investigated \cite{armstrong2023parafac2}.

Although retention information alone is often used to integrate univariate chromatographic separations data samples, such as those from Gas Chromatography - Flame Induction Detection (GC-FID) experiments, hyphenated methods that connect a multivariate detector in sequence to separation step are often used to improve the identification of each chemical component. This information can not only be used to assist in the interpretation of the data, but also for the alignment of the data, because the detector itself is intrinsically aligned through mass calibration in the case of a mass spectrometer and can be assumed not to vary significantly over the course of an experiment. While the raw data itself is multivariate, the chemical factors still suffer from their tendency to drift and sophisticated methods for integration are necessary to summarize the information as a peak table. Extensive research has been done into this type of data, owing to the popularity of High Performance Liquid Chromatography - Mass Spectrometry (LC-MS) data for the chemical analysis of bio-fluids in the Omics field. For Gas Chromatography - Mass Spectrometry (GC-MS) data, an elegant approach for analysis can be performed using PARAFAC2 \cite{kiers1999parafac2}, which includes a relaxed constraint for trilinearity across $M \times J \times N$ acquisitions (i.e., measurements (retention times)), mass-to-charge ratios (m/z) and samples (later: replicates in the final peak table) respectively. PARAFAC2 cannot handle all problems with the analysis of hyphenated chromatographic data in one step however; prior to analysis, regions of interest (ROIs) and their associated component numbers must also be selected \cite{baccolo2021untargeted,giebelhaus2022untargeted}.

Comparing strategies for the integration of chromatographic data (or ``pre-processing'' as it is commonly called), shows that the methodology used to extract the chemical features can drastically affect the results of a multivariate analysis~\cite{weggler2021unique}. And the best method is difficult to select for any problem when ``ground truth'' is unknown, as is common for all unsupervised learning.



This work suggests that an abstraction of the raw chromatographic data through a frequency domain representation can be used to subvert many common issues with chromatographic data analysis. This is done using the Fast Fourier Transform (FFT) on each sample within a designed experimental dataset collected using a Gas Chromatography Flame-Induction Detector (GC-FID) system followed by subsequent Analysis of Variance (ANOVA) - Simultaneous Component Analysis (ASCA)~\cite{smilde2005anova} that has been generalized to account for complex matrices. This technique recovers similar information to the more traditional peak table representation of the data, as provided by the authors of the original study, but requires an extra step to recover informative representations of the data in the time domain for the sake of interpreting the results. This is a simpler implementation of a method for analysis that is becoming widespread as an intermediate step in many popular algorithms ~\cite{schneide2023shift,schneide2024shift,yu2023parasias} that handle similar problems.


\subsection{ANOVA-Simultaneous Component Analysis}

The following description of ASCA is summarized as per \cite{camacho2023permutation}. A matrix of experimental observations, $\mathbf{X} (N \times M)$ can be factorized according to its encoded factor matrix, $\mathbf{D} (N \times F)$ according to:

\begin{equation}
    \mathbf{X} = \mathbf{D} \mathbf{\hat{\Theta}}^T + \mathbf{E}
\end{equation}

\noindent where $\mathbf{E}$ is an error matrix of the same dimensions as $\mathbf{X}$ that indicates the variance not accounted for by the model $\mathbf{D} \mathbf{\hat{\Theta}}^T$. $\mathbf{\hat{\Theta}}$ is typically calculated through a least-squares regression commonly referred to as a General Linear Model, which is equivalent to a standard ANOVA or ASCA based on averages when the experimental design is balanced, and provides some robustness against unbalanced experiments when it is not \cite{smilde2005anova}. The significance of each experimental factor, as a subset of the columns in $\mathbf{D}$ is typically performed using permutation testing:

\begin{equation}
p = \frac{\#\{\pi:F^\pi_{\nu_1,\nu_2} \geq F^{A}_{\nu_1,\nu_2}\} + 1}{\Pi+1}
\end{equation}

\noindent which for each permutation the calculated F-ratio ($F^\pi_{\nu_1,\nu_2}$) of the permutated data is compared against the nominal F-ratio of the un-permuted data ($F^{A}_{\nu_1,\nu_2}$) in order to simulate an empirical null distribution against which the significance of each experimental factor can be compared. The F-ratios are calculated as a function of the sum of squares (SSQ) of the data reconstructed according to some subset ($\mathbf{D}_A\mathbf{\hat{\Theta}}_A^T$) of $\mathbf{D}\mathbf{\hat{\Theta}}^T$ for an arbitrary factor, $A$:

\begin{equation}
    \mathbf{X}_A = \mathbf{C}_A\mathbf{\hat{\Theta}}^T_A,
\end{equation}

\noindent against the residual matrix $E$\footnote{Note that other F-ratios can be used, in particular in the presence of random/nested effects and interactions \cite{anderson2014permutational}.}, the calculated test statistic $F_A$ for factor $A$ normalized $\nu_1, \nu_2$ degrees of freedom:

 \begin{equation}\label{eq:sig_test}
    F^A_{\nu_1,\nu_2} = \frac{||\mathbf{X}_A||_F^2/\nu_1}{||\mathbf{E}||^2_F/\nu_2}
\end{equation}

\noindent where the squared Frobenius norm, denoted as $||\cdot||^2_F$, is used to indicate the sum of squares for the factor and residual matrices. For complex-valued data, the real-valued sum of squares are calculated through the following identity:

\begin{equation}\label{eq:ssq}
    SS_A = ||\mathbf{X}_A||^2_F = Tr(\mathbf{X}_A\mathbf{X}_A^H)
\end{equation}

\noindent where $\mathbf{X}^H_A$ is the \textit{Hermitian} (i.e. conjugate transpose) of $\mathbf{X}_A$ and must be explicitly calculated to ensure that a real-valued scalar is returned. Proof of this identity is widely-known, and can be derived by taking the matrix product of any complex data and its complex conjugate.

Altogether these reconstructed matrices are nominally orthogonal in the case of a balanced design, and can be represented as summation of their contributions to the overall matrix of experimental data $\mathbf{X}$:

\begin{equation}
    \mathbf{X} = \mathbf{1}\mu + \mathbf{X}_A + \mathbf{X}_B + \mathbf{X}_{AB} + \mathbf{E},
\end{equation}

\noindent and the face-splitting product of the columns representing the linear factors themselves (such as $A$ and $B$) can be used to calculate the interactions between the factors $AB$. $\mu$ is the $1\times N$ vector of variable-wise means.

\subsection{Fast Fourier Transforms}

A key characteristic of the pre-processing strategy proposed in this work involves a frequency domain representation of the raw experimental data. This is done using a Fast Fourier Transform (FFT), which is an algorithmic acceleration of the Discrete Fourier Transform representing the time-domain data as a series of complex coefficients. These coefficients encode the relative magnitude and offset of a series of sinusoidal components that sum to a least-squares representation of the original data. FFTs can be performed in $L$ modes \cite{smith1995handbook}, which may generalize to multi-modal separations and/or hyphenated data, but for the sake of demonstration only 1-dimensional transforms on univariate data will be considered. Take the 1-dimensional discrete Fourier transform of a univariate time signal, $x_m$, for $m \in \{1...M\}$ acquisitions in the time domain:

\begin{equation}\label{eq:fft}
    x_k = \sum_{m=0}^{M-1} x_m e^{-i \frac{2\pi}{M} km}\textnormal{, for } k \in \{1...M/2\}
\end{equation}

\noindent where the right hand side of Equation \ref{eq:fft} transforms the data into $k$ frequencies up to $M/2$, or half the total number of acquisitions. The relationship is linear, and can be inverted:

\begin{equation}\label{eq:ifft}
    x_m = \frac{1}{M} \sum_{k=0}^{M-1} x_k e^{i \frac{2\pi}{M} kn}
\end{equation}

Note that the conventions for use of the normalization constant $\frac{1}{M}$ for the forward or inverse transforms vary by definition, since for equi-spaced data the transform is \textit{orthogonal}, but not \textit{orthonormal} to itself \cite{armstrong2023direct}.

\begin{figure}[tbh!]
\centering
\begin{tikzpicture}

\def\radius{5}
\def\angle{45} 

\pgfmathsetmacro\x{\radius * cos(\angle)}
\pgfmathsetmacro\y{\radius * sin(\angle)}

\draw[very thick, ->] (-1.5,0) -- (\radius+0.5,0) node[right] {\LARGE Re} node[midway, below] {$|n| = a$};
\draw[very thick, ->] (0,-1.5) -- (0,\radius+0.5) node[above] {\LARGE Im};

\coordinate (A) at (\x,\y);
\draw[very thick, fill] (A) circle [radius=0.05] node[above right] {$|z| = \sqrt{a^2 + b^2}$};

\coordinate (O) at (0,0);
\draw[very thick, fill] (O) circle [radius=0.05] node[below left] {$0$};

\draw[-, very thick] (O) -- (A) node[midway,above, sloped]{};

\draw[very thick, dashed] (\radius,0) arc[start angle=0, end angle=90, radius=\radius];

\draw[thick, dashed] (A) -- (\x,0) node[midway, right] {$b$};
\draw[thick, dashed] (A) -- (0,\y) node[midway, above] {$a$};

\end{tikzpicture}

\caption{Calculation of the magnitude of the complex coefficients as the euclidean distance from $0 + 0i$, as it relates to the implicit magnitude of real-valued data ($|n|=a$).}
\label{fig:imag}
\end{figure}
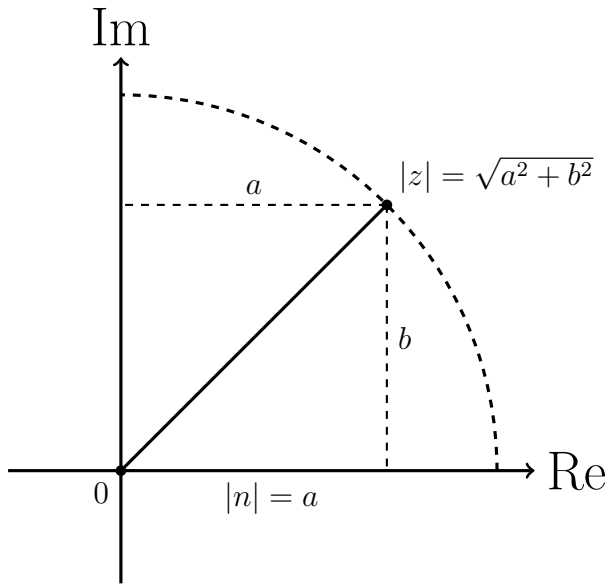

In this case, for a series of time-series data representing $N$ chromatographic samples, $x_m$, is indexed row-wise and each are transformed into the frequency domain via Equation \ref{eq:fft}. An ASCA analysis is then performed on the matrix of the complex coefficients - where statistical significance testing is performed on the magnitude of each complex coefficient according to Figure \ref{fig:imag}, and Equation \ref{eq:ssq}.

Significance testing follows as normally, and the complex characteristics of the data are preserved for the subsequent visualization step. The loadings of the reconstructed matrices for each experimental factor are projected using summed reconstruction and error terms to calculate the scores for the ASCA model according to an arbitrary factor $A$:
\begin{align}\label{eq:asca}
    \mathbf{T}_A\mathbf{P}_A^H = \mathbf{X}_A \\
    \mathbf{T}^*_A = (\mathbf{X}_A + \mathbf{E})\mathbf{P}_A
\end{align}
\noindent where the scores and loadings matrices $\mathbf{T}_A$ and $\mathbf{P}_A$ are respectively $N \times R$, and $M \times R$ complex matrices, with $\mathbf{P}_A^H$ representing the Hermitian of the loadings matrix $\mathbf{P}_A$. In the frequency domain, the loadings $\mathbf{P}_A$ may not be particularly easy to interpret as an abstraction of the original data in the time series. But the loadings can be transformed back into the time domain without sacrificing the least-squares solution to the GLM-like problem as in Equation \ref{eq:GLM}. Consider $\mathbf{X}_m$ as the time-domain representation of a designed experimental dataset with one significant factor, and $\mathbf{X}_k$ as the corresponding frequency domain representation. Equations \ref{eq:fft} and \ref{eq:ifft} can be represented by the $M\times M$ linear transformation $\mathbf{F}$ for the Discrete Fourier Transform (DFT), and $\frac{1}{M}\mathbf{F}^{-H} = \mathbf{F}^{-1}$ for the corresponding inverse scaled by the number of observations, $M$. It follows that:

\begin{align}\label{eq:least_squares}
    \mathbf{X}_k = \mathbf{X}_m\mathbf{F}\\
    \mathbf{X}_m = \mathbf{X}_k\mathbf{F}^{-1}\\
    \mathbf{X}_k = \mathbf{T}\mathbf{P}^H + \mathbf{E}\\
    \mathbf{X}_m = \left(\mathbf{T}\mathbf{P}^H + \mathbf{E}\right)\mathbf{F}^{-1}\\
    \mathbf{X}_m = \mathbf{T}\left(\mathbf{P}^H\mathbf{F}^{-1}\right) + \mathbf{E}\mathbf{F}^{-1}
\end{align}

And so the transform of the loadings from the frequency domain encompass similar variance as a function of the linear transformation $\mathbf{F}$, and can be interpreted following the inverse transform. Additionally, the reconstructions of $\mathbf{X}$ according to the different experimental factors may be transformed back to the time domain for an inspection of the different peaks responsible for affecting the level separations. Inspecting the equations shows that the relationship is \textit{least squares} in the frequency domain, which is not to say the same relationship holds in the time domain. Following the reconstruction of the data according to the \textit{treatment} factor in Figure \ref{fig:freq_treat}, the varying width of the lines illustrate this in regards to peak drift in the time domain.

\subsection{Experimental Data and Methodology}

Data from a study measuring the changes in chemical profiles of the red flour beetle \textit{Tribolium castaneum}, as a function of immune stimulation. The motivation for the original study \cite{lo2023immune} was to investigate how the insects communicate via chemical signalling in response to environmental stressors. One of the two experimental datasets for this design was used in this example, which analysed stink glad secretions 24h and 72h following the absence of any exposure to environmental stressors, exposure to physical wounding via sterile PBS injection, and exposure to inactive \textit{Bacilius thuringiensis bv. tenebrionis}. Each sample represents the stink gland secretions of one individual, collected through chemical extraction and analysed using GC-FID. In addition to the sex of the individuals being a factor in the design, the order in which the samples were run were considered as an additional experimental variable which is presumed not to be significant. The contributions of each factor to the overall variance in the model is summarized as:

\begin{equation}\label{eq:GLM}
    \mathbf{X} = \mathbf{T} + \mathbf{R} + \mathbf{S} + \mathbf{O} + \mathbf{E}
\end{equation}

\noindent which encompasses $\mathbf{T}$, $\mathbf{R}$, $\mathbf{S}$, and $\mathbf{O}$ as the reconstructed matrices for ``Time'', ``Treatment'', ``Sex'', and ``Order'' respectively. 

All experimental data were analyzed using a modified variant of the MEDA Toolbox (to handle complex data) on MATLAB 2024a. The calculations were performed on system running Ubuntu 22.04.4 LTS with an Intel i9-1400K 32 core CPU, two parallel NVIDIA GeForce RTX 4070 GPUs and 128 GB of RAM. Experimental meta-data were converted to numerical format using Python. All code used in the analysis are available in the repository \url{https://github.com/CoDaSLab/fftasca_gcfid}.

\section{Results and Discussion}

\subsection{Analysis of a synthetic dataset}

A synthetic GC-FID dataset was made using an approach to simulate multivariate relationships \cite{Camacho2016} for one significant linear factor. The output of this routine was used to inform the amplitude of 5 Gaussian peaks, with an additional 5 whose amplitudes were randomly generated. Across each simulated samples' 5000 acquisitions, the Gaussian peak widths of $4\sigma$ were nominally sampled 20 times, which is typical for such data. Random peak drift, or ``jitter'' on the order of 0 to 50 acquisitions was generated while preserving the relative peak order, and at each instance an parallel GLM analysis was performed on the raw data (time), versus a similar analysis performed following an FFT transformation (frequency). At each level, a randomly generated synthetic dataset was created for $n=10$ replicates and the observed \textit{p}-value was converted to a \textit{z} score and compared against the \textit{z} score known from the synthetic data, assuming a latent normal distribution.

\begin{figure}[hbt!]
    \centering
    \includegraphics[width=0.9\linewidth]{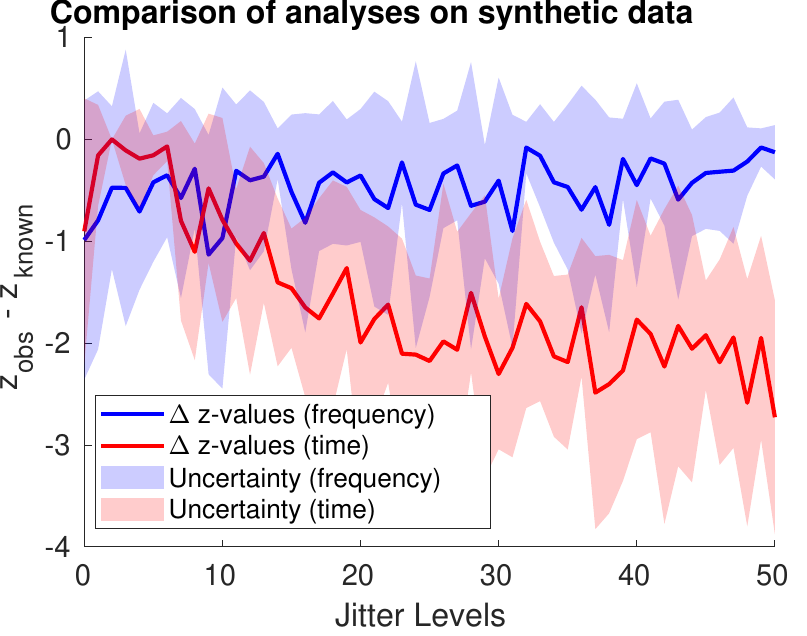}
    \caption{As shown by the results of this analysis, as the jitter in the data increases, the hypothesis testing step in parGLM analysis using the time-domain data becomes much less sensitive. However, following pre-processing using an FFT analysis the results are much more consistent.}
    \label{fig:compare}
\end{figure}

The results in Figure \ref{fig:compare} suggest that the evidence for significance of the one linear factor is better preserved as a function of increasing jitter, versus the time-series analysis. It was observed that when the order of the peaks changed, this relationship was no longer valid, which indicates that this method is still not appropriate for truly heterogeneous data. Nonetheless, the results suggest its applicability to well-controlled experimental data with relatively minor drift in retention times that do not affect the overall topology of the data.

\subsection{Analysis of real data}

The results of several ASCA models are shown below in Table \ref{tb:results}, using the raw chromatographic signals from \cite{lo2023immune}, and the corresponding MetaboLights repository (\url{https://www.ebi.ac.uk/metabolights/editor/MTBLS2277}). The peak tables provided as part of the original study were mean-centred. The effect of missing values in the peak table data were examined by performing the ASCA analysis on the data with and without a permutational conditional mean replacement strategy for minimizing the effect of missing data on measuring the effect of different experimental factors. In short, for each permutation, the missing values are replaced with the apparent \textit{cell mean} as the shifted data presents relative to the experimental design matrix which is kept constant. All zeros were presumed to be missing values.

For each GLM analysis, all binary interacting terms were examined before being ``trimmed'' in a subsequent analysis where only the previously indicated terms were included in the model. Interacting terms whose linear factors were not indicated as significant were not considered, except for interaction between Time and Treatment which is close to the cutoff for $\alpha = 0.05$ and further demonstrates consistency between the peak table analysis accounting for missing data, and the FFT-ASCA results.

\begin{table} 
\begin{tabular}{llllllll}
\textbf{Peak Table w/ zeros} & SumSq & PercSumSq & df & MeanSq & F & Pvalue \\ 
 \hline 
Mean & 2.2e+08 & 67.1 & 1 & 2.2e+08 & -- & -- \\ 
Time-1 & 7.53e+05 & 0.229 & 1 & 7.53e+05 & 0.747 & 0.412 \\ 
Treatment-2 & 9.45e+06 & 2.88 & 2 & 4.73e+06 & 4.69 & \textbf{0.00999} \\ 
Sex-3 & 1e+07 & 3.06 & 1 & 1e+07 & 9.96 & \textbf{0.002} \\ 
Order-4 & 5.26e+06 & 1.6 & 7 & 7.51e+05 & 0.745 & 0.675 \\ 
Interaction 1-2 & 4.19e+06 & 1.27 & 2 & 2.09e+06 & 2.08 & 0.113 \\ 
Residuals & 7.86e+07 & 23.9 & 78 & 1.01e+06 & -- & -- \\ 
Total & 3.29e+08 & 100 & 92 & 3.57e+06 & -- & -- \\ 

 \\
 \textbf{Peak Table pCMR}& SumSq & PercSumSq & df & MeanSq & F & Pvalue \\ 
 \hline 
Mean & 5.2e+03 & 68.6 & 1 & 5.2e+03 & -- & -- \\ 
Time-1 & 80.9 & 1.07 & 1 & 80.9 & 3.67 & \textbf{0.022} \\ 
Treatment-2 & 115 & 1.52 & 2 & 57.5 & 2.6 & \textbf{0.018} \\ 
Sex-3 & 180 & 2.38 & 1 & 180 & 8.17 & \textbf{0.000999} \\ 
Order-4 & 161 & 2.13 & 7 & 23 & 1.04 & 0.381 \\ 
Interaction 1-2 & 92.1 & 1.22 & 2 & 46.1 & 2.09 & 0.0569 \\ 
Residuals & 1.72e+03 & 22.7 & 78 & 22.1 & -- & -- \\ 
Total & 7.58e+03 & 100 & 92 & 82.4 & -- & -- \\ 

 \\ 
 \textbf{FFT, mean-centred}& SumSq & PercSumSq & df & MeanSq & F & Pvalue \\ 
 \hline 
Mean & 5.84e+14 & 5.09 & 1 & 5.84e+14 & -- & -- \\ 
Time-1 & 1.54e+15 & 13.4 & 1 & 1.54e+15 & 19.5 & \textbf{0.000999} \\ 
Treatment-2 & 7.79e+14 & 6.78 & 2 & 3.89e+14 & 4.92 & \textbf{0.00599} \\ 
Sex-3 & 8.7e+14 & 7.57 & 1 & 8.7e+14 & 11 & \textbf{0.00599} \\ 
Order-4 & 7.13e+14 & 6.2 & 7 & 1.02e+14 & 1.29 & 0.25 \\ 
Interaction 1-2 & 4.51e+14 & 3.93 & 2 & 2.26e+14 & 2.85 & 0.0569 \\ 
Residuals & 6.49e+15 & 56.5 & 82 & 7.92e+13 & -- & -- \\ 
Total & 1.15e+16 & 100 & 96 & 1.2e+14 & -- & -- \\ 
\end{tabular} 
 \\

\caption{Results of GLM analyses of the experimental factors versus the different representations of the multivariate data. Beginning with the un-altered peak table, followed by the zero-handling GLM with permutational conditional mean replacement (pCMR) \cite{pCMR}, and then the mean-centred frequency domain GLM analysis which contains no missing information. P-values below 0.05 are shown in bold.}
\label{tb:results}

\end{table}

From the results of Table \ref{tb:results} it is clear from the apparent significance of the experimental factors, that the presence of zeros has a profound effect on the resulting interpretation of the data. In the peak table representation, failure to account for zeros neglects to demonstrate the significance of the time factor, and suppresses the relative weight of the interacting term. Because the peak table representation relies on an intermediate step to integrate the peaks, it is impossible to know which zeros are truly absent from the data, and which zeros have failed to be integrated by the software. However it is worth noting that a similar treatment in the time and frequency domains yields similar results when the zeros are accounted for by the permutational cell mean replacement routine. 

Figures \ref{fig:figure1} and \ref{fig:figure2} demonstrate that the scores from the analysis appear to be more consistent using FFT-ASCA than by using ASCA on the peak table information alone, and the discrete peak loadings in Figure \ref{fig:figure5} show some similarity to the continuous, inverse transformed loadings in Figure \ref{fig:figure6}. This further highlights that pCMR is currently only capable of handling issues with significance testing, and that visualization is still strongly influenced by the presence of missing data.  

\subsection{ASCA}

\begin{figure}[hbtp!]
    \centering
    \begin{subfigure}[b]{0.45\textwidth}
        \centering
        \includegraphics[width=\textwidth]{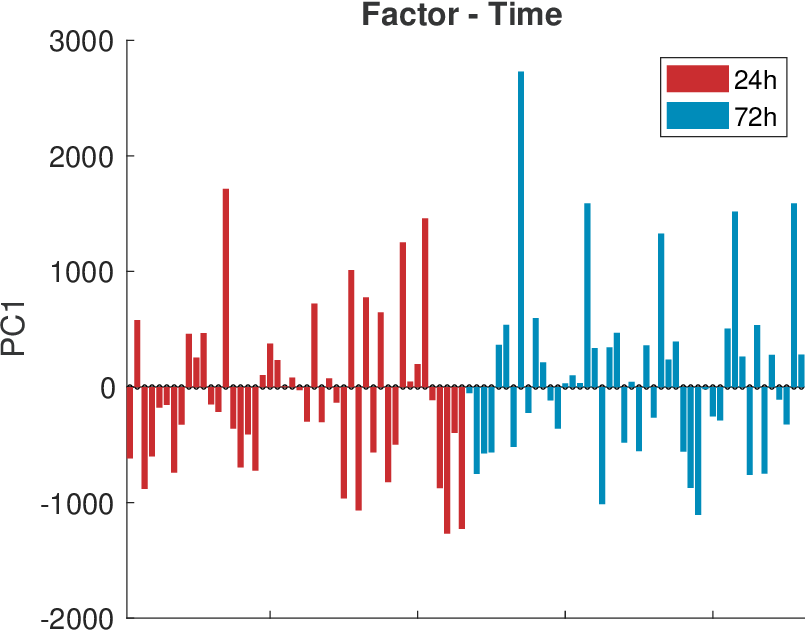}
        \caption{Peak table PCA scores of factor ``Time''.}
        \label{fig:figure1}
    \end{subfigure}
    \hfill
    \begin{subfigure}[b]{0.45\textwidth}
        \centering
        \includegraphics[width=\textwidth]{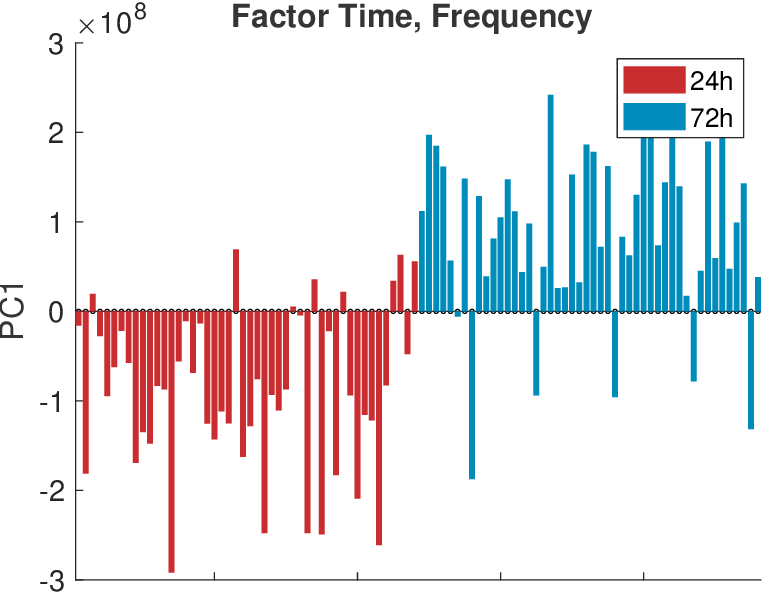}
        \caption{Real-valued PCA scores for factor ``Time'' from FFT transformed data. Mean-centred.}
        \label{fig:figure2}
    \end{subfigure}
    
    \vfill
    
    \begin{subfigure}[b]{0.45\textwidth}
        \centering
        \includegraphics[width=\textwidth]{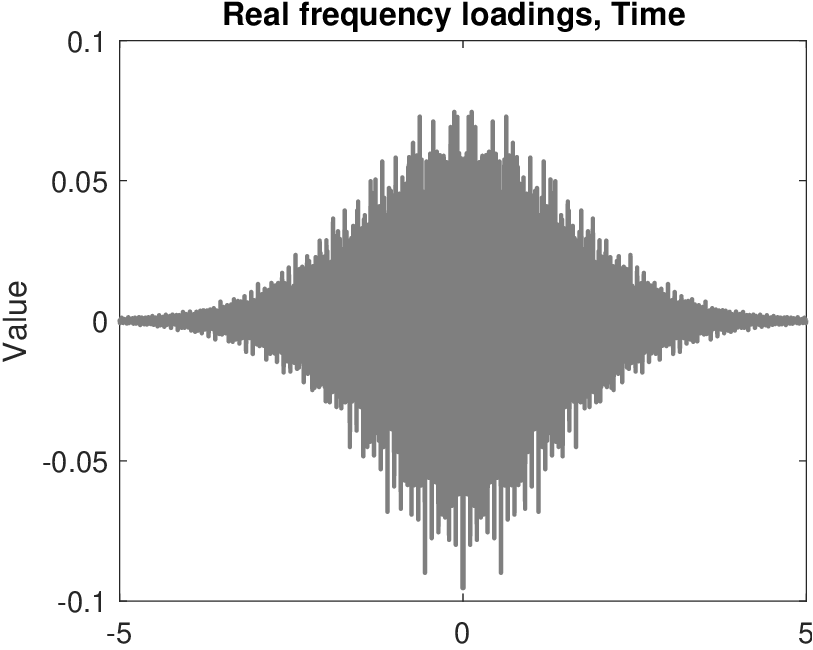}
        \caption{Loadings for factor "Time" from the peak table representation of the data. 26 targeted metabolites were included in the analysis.}
        \label{fig:figure3}
    \end{subfigure}
    \hfill
    \begin{subfigure}[b]{0.45\textwidth}
        \centering
        \includegraphics[width=\textwidth]{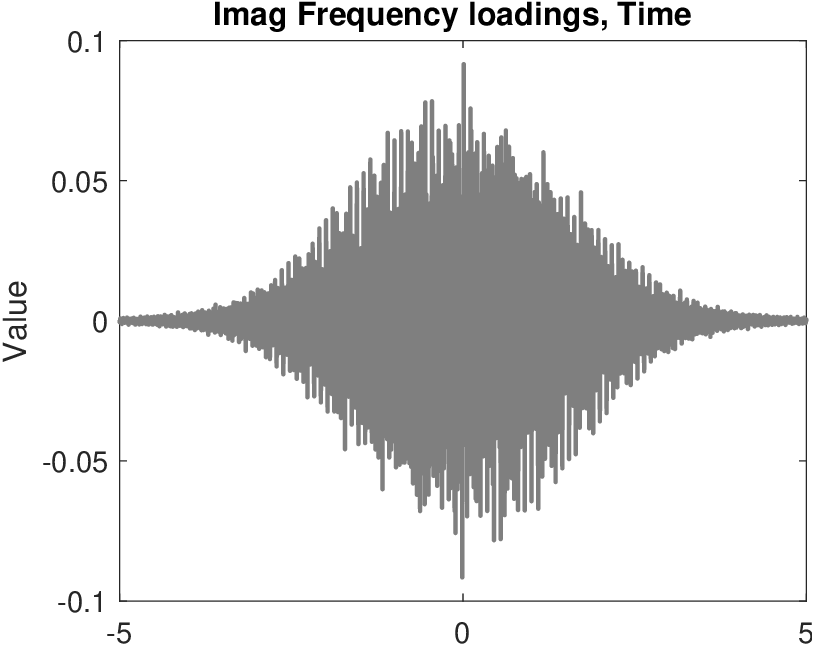}
        \caption{Complex-valued loadings in the frequency domain for factor ``Time''.}
        \label{fig:figure4}
    \end{subfigure}
    \vfill

\begin{subfigure}[b]{0.45\textwidth}
        \centering
        \includegraphics[width=\textwidth]{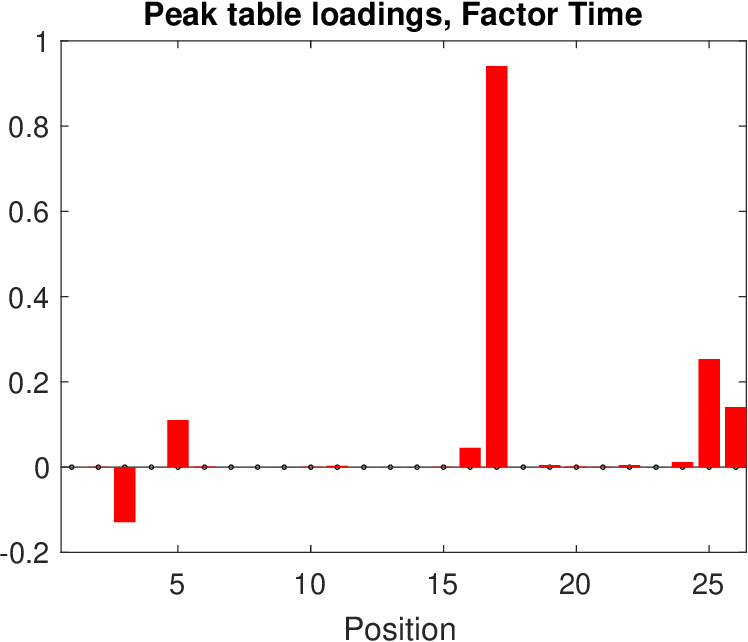}
        \caption{Real-valued loadings in the time domain for factor ``Time''.}
        \label{fig:figure5}
    \end{subfigure}
    \hfill
    \begin{subfigure}[b]{0.45\textwidth}
        \centering
        \includegraphics[width=\textwidth]{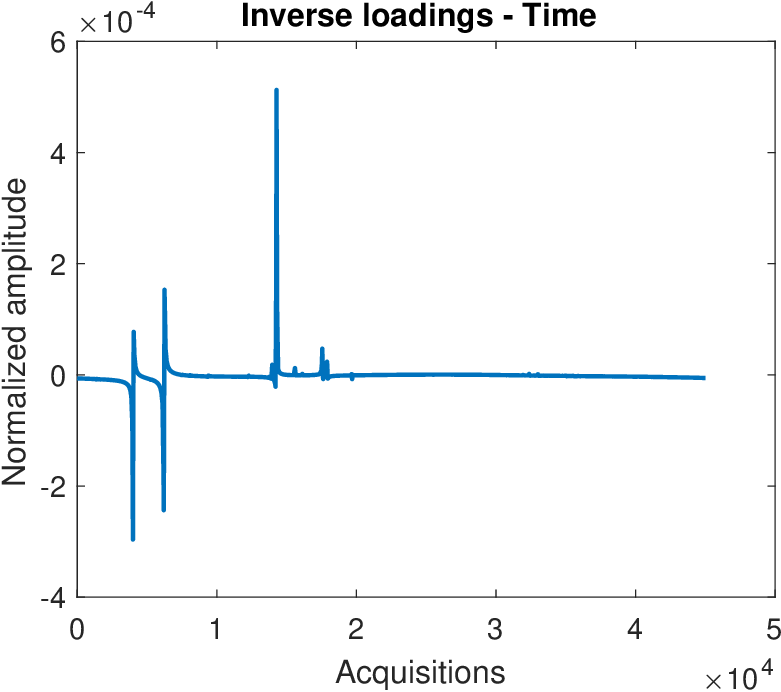}
        \caption{Real-valued loadings in the time domain for factor ``Time''.}
        \label{fig:figure6}
    \end{subfigure}
    \caption{ASCA results for the peak table analysis, as well as the frequency-domain analysis. The real-valued scores in the frequency domain are more consistent (Figure \ref{fig:figure2}, versus Figure \ref{fig:figure1}), but the loadings themselves are less interpretable (Figures \ref{fig:figure3} and \ref{fig:figure4}). They can however be transformed back into the time domain, as shown in Figure \ref{fig:figure6}.}
    \label{fig:subfigures}
\end{figure}

From Equation \ref{eq:asca} the following \textit{post-hoc} visualizations of the data were performed in agreement with significance testing results. The factor ``Time'' was indicated as significant only after correcting for missing values in the peak table data, but this routine does not affect the factorization and visualization by ASCA (only the inference), which does not yield especially interpretable results (see Figure \ref{fig:figure1}). 

\begin{figure}[hbt!]
    \centering
    \includegraphics[width=0.9\linewidth]{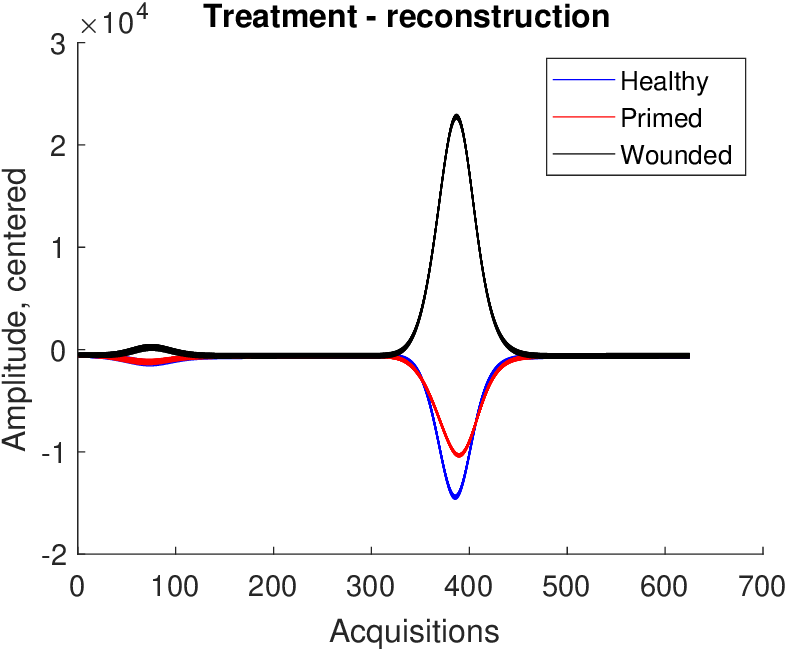}
    \caption{Time-domain reconstruction of factor ``Treatment'' where the levels are clearly differentiated.}
    \label{fig:freq_treat}
\end{figure}

In addition to analyzing the loadings, the individual factor matrices themselves can be transformed back into the time domain for  inspection. This type of analysis shows that the GLM performed on the frequency domain, as an abstraction of the original, raw data still corresponds to real-valued chemical components. An example is shown in Figure \ref{fig:freq_treat}, where a single chemical factor with evidence of statistical significance in the frequency domain, is clearly differentiated according to its levels in the time domain.

\section{Conclusions}

FFT-ASCA analyses appear to subvert some common problems with pre-processing of chromatographic data. In the experimental dataset, the presence of missing values in the peak table was shown to have a strong effect on the resulting interpretation. A transformation of the raw data itself however to a series of fully-populated Fourier coefficients offers similar insight into the analysis following missing value correction, and for a similar level of interpretability. While simple, this method offers new ways of analyzing chromatographic data, in a manner that relies on fewer meta-parameters for feature engineering and extraction prior to the analysis step. Although this technique has not been definitely proven to out-perform traditional methods for analyzing chromatographic data in all cases, this demonstration is representative of a number of issues faced by laboratory scientists handling and interpreting chromatographic data. GC-FID is not commonly used for metabolomics-type work, but this type of methodology is easily generalized to hyphenated and/or multi-modal separations using similar extensions of multivariate statistics to tensor analysis that have already been widely discussed in literature \cite{koleini2023complementary}. 

\section{Acknowledgements}

This work was supported by the Agencia Estatal de Investigación in Spain for call no. MCIN/AEI/10.13039/501100011033 and grant no. PID2023-1523010B-IOO (MuSTARD). Michael Sorochan Armstrong has received funding from the European Union's Horizon Europe research and innovation programme under the Marie Skłodowska-Curie grant agreement no. 101106986 (MAHOD).

 \bibliographystyle{elsarticle-num} 
 \bibliography{cas-refs}

\begin{thebibliography}{10}
\expandafter\ifx\csname url\endcsname\relax
  \def\url#1{\texttt{#1}}\fi
\expandafter\ifx\csname urlprefix\endcsname\relax\def\urlprefix{URL }\fi
\expandafter\ifx\csname href\endcsname\relax
  \def\href#1#2{#2} \def\path#1{#1}\fi

\bibitem{christensen2005chromatographic}
J.~H. Christensen, J.~Mortensen, A.~B. Hansen, O.~Andersen, Chromatographic
  preprocessing of gc--ms data for analysis of complex chemical mixtures,
  Journal of Chromatography A 1062~(1) (2005) 113--123.

\bibitem{armstrong2023parafac2}
M.~D.~S. Armstrong, J.~L. Hinrich, A.~P. de~la Mata, J.~J. Harynuk,
  Parafac2$\times$ n: coupled decomposition of multi-modal data with drift in n
  modes, Analytica Chimica Acta 1249 (2023) 340909.

\bibitem{kiers1999parafac2}
H.~A. Kiers, J.~M. Ten~Berge, R.~Bro, Parafac2—part i. a direct fitting
  algorithm for the parafac2 model, Journal of Chemometrics: A Journal of the
  Chemometrics Society 13~(3-4) (1999) 275--294.

\bibitem{baccolo2021untargeted}
G.~Baccolo, B.~Quintanilla-Casas, S.~Vichi, D.~Augustijn, R.~Bro, From
  untargeted chemical profiling to peak tables--a fully automated ai driven
  approach to untargeted gc-ms, TrAC Trends in Analytical Chemistry 145 (2021)
  116451.

\bibitem{giebelhaus2022untargeted}
R.~T. Giebelhaus, M.~D.~S. Armstrong, A.~P. de~la Mata, J.~J. Harynuk,
  Untargeted region of interest selection for gas chromatography--mass
  spectrometry data using a pseudo f-ratio moving window, Journal of
  Chromatography A 1682 (2022) 463499.

\bibitem{weggler2021unique}
B.~A. Weggler, L.~M. Dubois, N.~Gawlitta, T.~Gr{\"o}ger, J.~Moncur,
  L.~Mondello, S.~Reichenbach, P.~Tranchida, Z.~Zhao, R.~Zimmermann, et~al., A
  unique data analysis framework and open source benchmark data set for the
  analysis of comprehensive two-dimensional gas chromatography software,
  Journal of Chromatography A 1635 (2021) 461721.

\bibitem{smilde2005anova}
A.~K. Smilde, J.~J. Jansen, H.~C. Hoefsloot, R.-J.~A. Lamers, J.~Van Der~Greef,
  M.~E. Timmerman, Anova-simultaneous component analysis (asca): a new tool for
  analyzing designed metabolomics data, Bioinformatics 21~(13) (2005)
  3043--3048.

\bibitem{schneide2023shift}
P.-A. Schneide, R.~Bro, N.~B. Gallagher, Shift-invariant tri-linearity—a new
  model for resolving untargeted gas chromatography coupled mass spectrometry
  data, Journal of Chemometrics 37~(8) (2023) e3501.

\bibitem{schneide2024shift}
P.-A. Schneide, N.~B. Gallagher, R.~Bro, Shift invariant soft trilinearity:
  Modelling shifts and shape changes in gas-chromatography coupled mass
  spectrometry., Chemometrics and Intelligent Laboratory Systems (2024) 105155.

\bibitem{yu2023parasias}
H.~Yu, R.~Bro, N.~B. Gallagher, Parasias: A new method for analyzing
  higher-order tensors with shifting profiles, Analytica Chimica Acta 1238
  (2023) 339848.

\bibitem{camacho2023permutation}
J.~Camacho, C.~D{\'\i}az, P.~S{\'a}nchez-Rovira, Permutation tests for asca in
  multivariate longitudinal intervention studies, Journal of Chemometrics
  37~(7) (2023) e3398.

\bibitem{anderson2014permutational}
M.~J. Anderson, Permutational multivariate analysis of variance (permanova),
  Wiley statsref: statistics reference online (2014) 1--15.

\bibitem{smith1995handbook}
W.~W. Smith, J.~Smith, Handbook of real-time fast Fourier transforms, IEEE New
  York, 1995.

\bibitem{armstrong2023direct}
M.~S. Armstrong, J.~C. P{\'e}rez-Gir{\'o}n, J.~Camacho, R.~Zamora, A direct
  solution to the interpolative inverse non-uniform fast fourier transform
  problem, for spectral analyses of non-equidistant time-series data, arXiv
  preprint arXiv:2310.15310 (2023).

\bibitem{lo2023immune}
L.~K. Lo, L.~J. Tewes, B.~Milutinovi{\'c}, C.~M{\"u}ller, J.~Kurtz, Immune
  stimulation via wounding alters chemical profiles of adult tribolium
  castaneum, Journal of chemical ecology 49~(1) (2023) 46--58.

\bibitem{Camacho2016}
J.~Camacho,
  \href{http://www.sciencedirect.com/science/article/pii/S0169743916305020}{{O}n
  the {G}eneration of {R}andom {M}ultivariate {D}ata}, Chemometrics and
  Intelligent Laboratory Systems 160 (2017) 40 -- 51.
\newblock \href
  {https://doi.org/http://dx.doi.org/10.1016/j.chemolab.2016.11.013}
  {\path{doi:http://dx.doi.org/10.1016/j.chemolab.2016.11.013}}.
\newline\urlprefix\url{http://www.sciencedirect.com/science/article/pii/S0169743916305020}

\bibitem{pCMR}
O.~Merchanskaya, M.~Sorochan~Armstrong, C.~Gómez-Llorente, P.~Ferrer,
  S.~Fernandez-Gonzalez, M.~Perez-Cruz, M.~Gómez-Roig, J.~Camacho,
  Considerations for missing data, outliers and transformations in permutation
  testing for anova, asca(+) and related factorizations, Submitted to
  Chemometrics and Intelligent Laboratory Systems (2024).

\bibitem{koleini2023complementary}
F.~Koleini, S.~Hugelier, M.~A. Lakeh, H.~Abdollahi, J.~Camacho, P.~J.
  Gemperline, On the complementary nature of anova simultaneous component
  analysis (asca+) and tucker3 tensor decompositions on designed multi-way
  datasets, Journal of Chemometrics 37~(11) (2023) e3514.

\end{thebibliography}





\end{document}